# NUMERICAL SIMULATIONS OF RUNAWAY ELECTRON GENERATION IN PRESSURIZED GASES

D. Levko, S. Yatom, V. Vekselman, J. Z. Gleizer, V. Tz. Gurovich, and Ya. E. Krasik

Department of Physics, Technion, 32000 Haifa, Israel

The results of a numerical simulation of the generation of runaway electrons in pressurized nitrogen and helium gases are presented. It was shown that runaway electrons generation occurs in two stages. In the first stage, runaway electrons are composed of the electrons emitted by the cathode and produced in gas ionization in the vicinity of the cathode. This stage is terminated with the formation of the virtual cathode, which becomes the primary source of runaway electrons in the second stage. Also, it was shown that runaway electrons current is limited by both the shielding of the field emission by the space charge of the emitted electrons and the formation of a virtual cathode. In addition, the influence of the initial conditions, such as voltage rise time and amplitude, gas pressure, and the type of gas, on the processes that accompany runaway electrons generation is presented.

## I. INTRODUCTION

Today, nanosecond pulsed high-voltage (HV) and high-current discharges in pressurized gases are applied in various applications: plasma-assisted combustion,[1,2] pulsed gaseous lasers,[3] and the generation of electron beams and x-rays.[4,5] Nevertheless, nowadays there is no model, including the commonly accepted Townsend model, that can describe the fast avalanching processes in the over-voltage discharge gaps accompanying such a type of discharge. Currently, runaway electrons (RAE) as a source of background electrons,[6-8] which play a major role in the discharge formation, are used to explain the evolution of such a type of gas discharge. Here, RAE are the electrons that acquire more energy during their acceleration by an electric field along a mean free path than they lose in inelastic collisions. RAE efficiently generate secondary ions and electrons that produce gas pre-ionization during their propagation toward the anode. Also, these secondary electrons participate in ionization processes, forming plasma that could result in cathode-anode (CA) gap shorting.

RAE generation has been studied using numerical simulation in many research studies (see, for instance, Refs. 7-12). One of the first numerical simulations of RAE generation in pressurized gases was carried out by Kunhardt et al.[9] where electron avalanche propagation in an external electric field with $E \geq 6 \times 10^4$ V/cm was studied, and the electron energy distribution (EEDF) versus the distance from the cathode and avalanche dimensions was determined. However, these simulations did not consider such important processes as the electron field emission from the cathode and the influence of secondary electrons and ions space charge on the external electric field.

In numerical simulations carried out by Yakovlenko et al. (see Ref. 8), RAE generation was studied for different electrode configurations (planar or cylindrical geometry), types of background gas (air, nitrogen, helium, neon, argon and $SF_6$), and accelerating voltage waveforms. The simulations were carried out using one-dimensional Particle-in-Cell (1D PIC) numerical simulations for planar electrodes' geometry. These simulations do not consider the changes in electric field distribution caused by the secondary electrons and ions space charge. In addition, the process of electrons field emission (FE) was not accounted for in these simulations. It was shown that the breakdown of over-voltage gas-filled gaps could be described by the Townsend model when the distance $d_{ca}$ between the CA electrodes exceeds some critical length, $l_{cr}$. In this case, the results of numerical simulations showed that the maximum of the EEDF at the anode corresponds to electron energies $\varepsilon^* \ll e\varphi_c$, where $\varphi_c$ is the cathode potential and $e$ is the electron charge. When $l_{cr} > d_{ca}$, the Townsend model failed to describe the breakdown formation. In this case, the majority of the electrons was found to be accelerated continuously, forming RAE, and the maximum of EEDF was obtained at $\varepsilon^* \approx e\varphi_c$. It was shown that the volume discharge in a non-uniform electric field is developed by background electrons formed in the process of gas ionization by RAE. In addition, these simulations showed that RAE generation occurred only near the cathode where one obtains the largest electric field.

G. Mesyats et al.[4] carried out numerical simulations of electrons generation in a diode using PIC code, considering electric field enhancement at the cathode surface micro-protrusions and shielding of the external electric field by the space charge of FE electrons. However, these simulations did not account for inelastic collisions and the scattering of electrons propagating toward the anode.

Comprehensive simulations of RAE generation in a non-uniform electric field for the initial stage (a few tens of picoseconds) of the gas discharge were carried out by V. Shklyaev, V. Ryzhov et al.[10] The developed numerical model considers the shielding of the FE and the self-consistent electric field distribution, i.e., the effect of the space charge of the secondary electrons and ions and emitted electrons was accounted for. The simulations showed that RAE generation occurred in the vicinity of the cathode during a few tens of picoseconds. The termination of RAE generation occurs when the electric field produced by the space charge of emitted electrons screens the external electric field to a significant extent. Also, the influence of the electric field enhancement by the cathode micro-protrusions and of the different work-functions of the cathode material on the RAE parameters was studied. It was shown that in the case of a low work-function,[11] a large amount of emitted electrons, which are already generated efficiently during the HV pre-pulse, produces the plasma (secondary electrons and ions) at distances of less than a few hundreds of microns from the cathode. This plasma becomes the source of RAE, which were emitted from its boundary during the main HV pulse. In addition, simulations showed the formation of the potential well that causes the capture of electrons with the energy smaller than the plasma potential. In the opposite case (i.e., large work-function), the amount of FE electrons decreases and plasma is not generated during the HV pre-pulse. Thus, RAE generation occurs only in the vicinity of the cathode. The results of these simulations of the initial stage of the gas discharge showed that RAE consist of both FE electrons and secondary electrons generated in the vicinity of the cathode.

In experimental research carried out during the last two decades (see, for instance, Refs. 4 and 5), it was found that RAE have several common features that do not depend on the initial conditions, such as gas type and pressure, CA gap, shape and material of the cathode, and the duration of the accelerating pulse. Namely, the duration of the RAE pulse does not excEEDF a few tens of picoseconds and RAE generation occurs only during the HV pulse rise time. At the present, the processes which are accompanied by RAE formation, and several other issues, such as the location(s) of RAE generation (in the cathode vicinity or in the entire volume of CA gap[7,8]) and the processes that terminates the RAE pulse duration (shielding of the electric field at the cathode by the emitted electrons or the transition from the FE to the explosive electron emission[12]), are not well understood and should be addressed further.

In this paper, the results of a 1D PIC simulation of RAE generation during the entire HV pulse in pressurized nitrogen ($N_2$) and helium (He) gases are presented. Namely, the main sources of RAE and the factors limiting the amplitude and duration of RAE are determined. The role of the initial conditions, such as the HV pulse rise time and amplitude and the gas type and pressure, on the RAE' parameters has been studied in detailed. The EEDF at the anode at different times of the HV pulse were obtained depending on the HV pulse parameters.

## II. NUMERICAL MODEL

In order to simulate the RAE generation in pressurized gases, a 1D PIC numerical code was developed for coaxial diode geometry with the wire cathode. Rough estimates showed that one can neglect the self-magnetic field of the diode current on the main processes governing the gas discharge and RAE generation. Indeed, the Lorentz force is ~$vB$, where $v$ is the electron velocity and $B = \mu_0 I/(2\pi r)$ is the magnetic field of the current carrying wire, $I$ is the wire current, and $r$ is the radius. The electric force is proportional to the electric field $E$, which in the vacuum case is $E = \varphi_c/[r \cdot \ln(r_A/r_c)]$, where $\varphi_c$ is the wire potential, and $r_c$ and $r_A$ are the cathode and anode radii, respectively. A simple analysis showed that the electric force plays the major role for electron energies $\varepsilon_e$<10keV in the case of the wire current $I \geq 1$kA and the diode geometry considered in this model. The simulations showed that in the vicinity of the cathode (<1mm) the value of $\varepsilon_e$<10keV. Therefore, here the electrons' propagation is governed mainly by the electric field. Let us note that the magnetic field does not change the electron collision frequency, which determines the rate of secondary plasma electron and ion generation in the vicinity of the wire. A 2D effect related to the finite time of electromagnetic wave propagation in the diode also can be neglected in the considered model. Indeed, the maximal potential difference between two ends of the wire being 1 cm length is ≤15kV, because of the finite time of the electromagnetic wave propagation for the sine-pulse of 500ps half a period and of 120kV in amplitude. Thus, one obtains a maximal axial electric field of ≤$1.5\times10^4$kV/cm whereas the radial electric field is ~$10^7$V/cm.

In the developed code, the radial distance-velocity phase space is divided into elementary cells with dimensions *dr* and *dv*. At each time interval, the system of equations for electron propagation in the local electric field was solved numerically.[13] Namely, in order to follow the energy conservation law, first new coordinates of electrons were calculated. Next, at the same time interval *dt*, the electrons' energy was calculated using new and old electron coordinates and the local electric field value that was calculated at the preceding time interval. The sequence of numerical simulations is shown in Fig. 1.

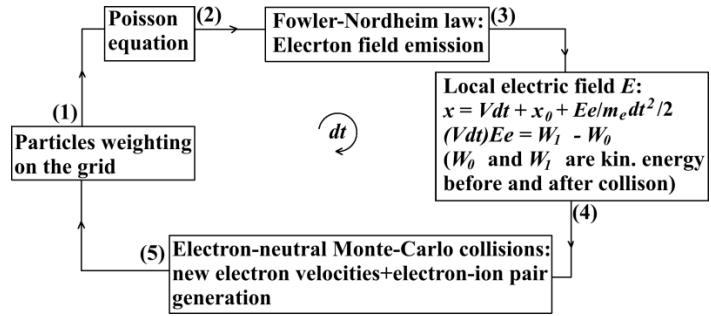

Fig.1. The sequence of numerical simulations.

The probability of collisions between neutrals and electrons in the cell was defined as:[14,15]

$$P = 1 - \exp[-\Delta r / \lambda(\varepsilon)]. \qquad (1)$$

Here, $\Delta r$ is the distance which electrons propagate during one time step, and $\lambda(\varepsilon)$ is the mean free path of electrons in $N_2$ gas. In the calculations of $\lambda$, the elastic scattering cross-section $\sigma_{el}$ of electrons by neutrals and the inelastic collisions cross-sections (ionization cross-section $\sigma_{ion}$ and excitation cross-section of the first electronic energy level of $N_2$, $\sigma_{ex}$) were accounted for, which results in the total mean free path:

$$\lambda(\varepsilon) = \frac{1}{N \cdot [\sigma_{el}(\varepsilon) + \sigma_{ion}(\varepsilon) + \sigma_{ex}(\varepsilon)]}. \tag{2}$$

Here $N$ is the density of neutrals. The NIST data base[16] was used for the ionization cross-section of $N_2$ gas molecules and elastic scattering. The excitation cross-sections were extrapolated for high energies (up to 200keV) using the values of cross-sections presented in Ref. 17. In addition, electron scattering forward and backward in both elastic and inelastic processes was considered.[14] The direction of the electron propagation after the collision was defined as $\eta = 1 - 2 \cdot \varepsilon_1 \cdot \left[ (1 + \varepsilon / \varepsilon_1)^\kappa - 1 \right] / \varepsilon$, where $\varepsilon_1$ = 1eV is the characteristic scattering energy,[18] and $\kappa$ is the arbitrary value, $0 \leq \kappa < 1$, which was used in the Monte Carlo sub-routine. In the case $\eta > 0$, the electrons do not change their propagation direction toward the anode. Otherwise, the electrons move toward the cathode after the collision.

In each process of molecule ionization, one electron-ion pair is generated. The newly generated secondary electron is added to the primary electrons. The velocities of the secondary electron and ion are assumed to be zero and their location is determined by the coordinate of the primary ionizing electron. In the model, the ions' motion was taken into account as well.

The electric field enhancement at the cathode surface was simulated using coaxial diode geometry, with a cathode having a length of 1cm and a radius of 3μm and an anode with a radius of 1cm. This diode configuration allows one to consider the cathode electric field enhancement that is typical for a blade-like cathode and a plane anode. This geometry is described adequately by the cylindrical CA geometry for an inter-electrode gap of 1cm. The radial potential distribution was calculated solving the Poisson equation at the beginning of each time step for new electrons and ions space charge distributions and new boundary conditions:

$$\frac{1}{r}\frac{d}{dr}\left(r\frac{d\varphi(r,t)}{dr}\right) = -\frac{\rho_i(r,t) - \rho_e(r,t)}{\varepsilon_0}. \tag{3}$$

Here $\varphi(r,t)$, $\rho_i(r,t)$ and $\rho_e(r,t)$ are the potential, ion and electron space charge densities, respectively, at the given time $t$ at a distance $r$ from the cathode. Here let us note that, in general, the variation in the diode current changes the diode voltage amplitude and waveform because of the finite internal impedance of the pulsed generator. In the developed model, this process, which is specific for each generator, was not accounted for. Namely, a simplified electrical circuit which allows one to obtain sine-like cathode potential was considered. Eq. (3) was solved with the cathode and anode boundary conditions for potential:

$$\varphi_c = -\varphi_0 \sin\left(\frac{2\pi t}{T}\right), \quad \varphi_a = 0. \tag{4}$$

Here, $\varphi_0$ is the maximal cathode potential whose value was varied in the simulations: 60kV, 120kV, 200kV and 280kV. The rise time of the cathode potential was determined as $T/4$, where $T$ is the period that was varied as 0.5ns, 1ns, and 2ns.

The electron emission from the cathode was governed by the FE described by the Fowler-Nordheim (FN) law.[19] At the beginning of each time step, a quantity $dn_{em}$ of electrons with zero velocity and zero (cathode) coordinates was added to the simulations according to the FN law. The value of $dn_{em}$ was determined as $dn_{em} = j_{FN} \cdot S \cdot dt/e$, where $j_{FN}$ is the electron current density and $S$ is the cathode surface area. The number $dn_{em}$ was added into the second space node to the electrons existing in this node, and the total space charge in this cell was used for the Poisson equation solution. Depending on the initial conditions, the time interval *dt* was varied in the range $10^{-15}$ - $10^{-14}$s allowing electrons to propagate $\Delta r \ll \lambda$ during *dt*. Several test simulations were carried out with different time and space steps to test the stability and accuracy of the obtained results.

In the simulations, a uniform FE of electrons from the cathode was considered. However, the cathodes used in the experiments have micro-protrusions whose distribution at the cathode surface, number, density and micro-protrusion apex dimensions are undefined variables that could be varied even during a single generator shot. Due to the large electric field enhancement at their apexes, these micro-protrusions could significantly change the parameters of FE and RAE.[11,20] However, one can consider two competing processes for FE of electrons from micro-protrusions. On the one hand, the smaller cross-sectional area of the micro-protrusions' apexes leads to a smaller quantity of emitted electrons. On the other hand, the number of electrons emitted from each micro-protrusion is significantly larger than from the wire for the same micro-protrusion apex area, due to a larger electric field enhancement at the micro-protrusion's apex for the same potential value. These two competitive processes allow one to decrease inaccuracy related to the concern of a uniform FE from cathode. Here let us note that Mesyats *et al.*[7] showed that the RAE generation is terminated when FE transfers to explosive electron emission. However, the results of the simulations showed that the FE current is $\sim 10^8$A/cm$^2$, which results in a time delay in the appearance of the explosive emission of $>10^{-7}$s for a wire that is 3μm in radius. Thus, a significantly smaller cathode radius is required to initiate the explosive emission for the HV pulse durations considered in the model.

**III. RESULTS AND DISCUSSION**

Simulations have shown that the vicinity of the cathode is the most important region for RAE generation. Therefore, the choice of electrode geometry could be important in numerical simulations. However, the comparison between electric field distributions for coaxial geometry and for geometry with a hyperbolic cathode and plane anode[21] (point-to-plane geometry) with the same initial conditions (length, radius and potential of the cathode, and CA gap) showed that there is no significant difference between the geometries. For instance, the comparison between geometries for $\varphi_c$=120kV and a CA gap of 10 mm showed that in the vicinity of the cathode the electric field is ~2 times larger for the point-to-plane geometry. In addition, for cylindrical geometry the electric field decreases to a critical electric field at a distance $r \approx 0.35$mm. At the same time, for point-to-plane geometry, the electric field decreases to a critical value at $r \approx 1.2$mm. Thus, one can conclude that qualitatively (and almost quantitatively) the processes governing RAE generation in the vicinity of the cathode are the same for both geometries.

At normal pressure ($P = 10^5$Pa) in $N_2$ gas the value of the critical electric field $E_{cr}$ that is necessary for electrons with the energies $\varepsilon_e > 40$eV to become RAE is $E_{cr} \geq 4.5 \times 10^5$V/cm.[7,8,22] Simulations showed that, in the present diode geometry, for $T = 2$ns one obtains electric field $E > E_{cr}$ at distances from the cathode up to $r < 14$μm at $t \geq 1.00$ps for $\varphi_0 = 120$kV, $t \geq 0.88$ps for $\varphi_0 = 200$kV, and $t \geq 0.65$ps for $\varphi_0 = 280$kV. However, RAE generation does not start at $E = E_{cr}$ because the first electron emission determined by FN law occurs only at the time when electric field at the cathode surface becomes $E_c \geq 2 \times 10^7$V/cm. For instance, for $\varphi_0 = 120$kV the first electron is emitted only at $t \approx 130$ps when $E_c = 2 \times 10^7$V/cm.

Here let us note that, in general, also background electrons with density $\leq 10^3$cm$^{-3}$ that exist naturally in gas[19] could become RAE.[23] However, the simulations showed that these electrons left the cathode region $r < 14\mu m$ at an earlier time when $E_c < E_{cr}$. Thus, these background electrons cannot be considered as the source of RAE.

Typical potential distributions and densities of electron and ion space charge at different times of the HV pulse ($T = 2$ns, $\varphi_0 = 120$kV) in the $N_2$ gas pre-filled ($P = 10^5$Pa) diode are shown in Fig. 2. One can see that at the beginning of the accelerating pulse ($t < 137$ps) the generated space charge of secondary electrons and ions does not change the distribution of the external electric field. Thus, at $t < 137$ps the electric field enhanced by the cathode geometry governs RAE generation. Further, when the space charge of electrons and ions increases, one obtains the formation of a potential hump [see Fig. 2(a), region denoted as KOK*] with a maximum potential temporal shift toward the cathode. Also, beginning at $t \geq 215$ps one obtains the formation of the virtual cathode (VC), i.e., the location with potential $\varphi_{VC} \approx \varphi_c$.[24] Depending on the initial conditions, the maximal value of the electric field, $E_{VC}$, toward the anode at the VC location reaches different values. Namely, for $T = 2$ns, $P = 10^5$Pa and $\varphi_0 = 120$kV, the value of $E_{VC} < E_{cr}$, which prevents RAE generation between the VC and the anode. Nevertheless, at $t > 260$ps the VC becomes the source of RAE because at that time the increase in the cathode potential and accumulated negative charge of secondary electrons at the VC location becomes sufficient to produce $E_{VC} > E_{cr}$. Also, one can see time- and space-redistribution of the VC negative charge, which leads to the VC shift toward the anode. Let us note that at time $t > 600$ps, the potential of the cathode exceeds that of the VC. The latter can be explained by the motion of secondary ions toward the cathode and the anode in the KO and OK* regions, respectively. The ions' motion toward the anode leads to partial compensation of the VC negative charge. In addition, the ions' motion toward the cathode leads to an increase in the electric field at the cathode what increases the $dn_{em}/dt$.

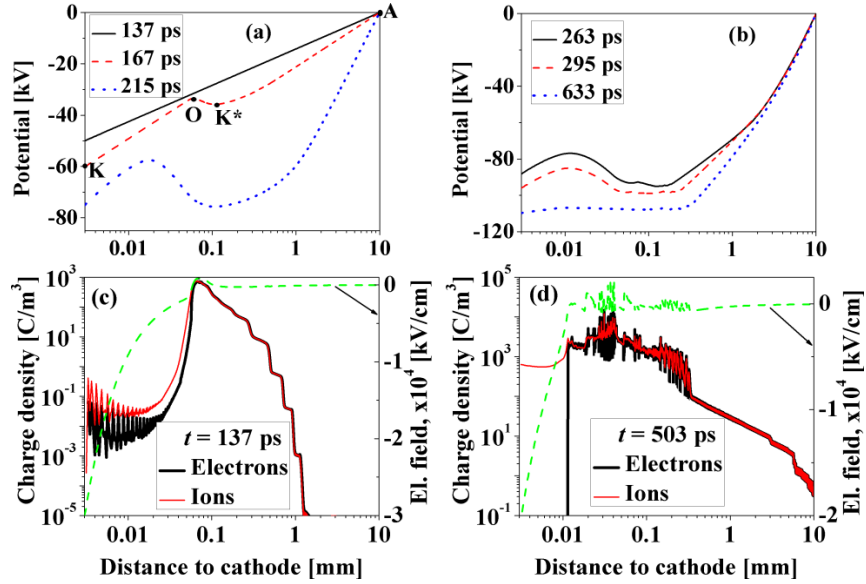

Fig. 2. (Color online) (a)-(b) Potential distribution in the CA gap and (c)-(d) distributions of charge densities of electrons and ions and electric field at different times of the accelerating pulse; $T = 2$ns, $\varphi_0 = 120$kV, $P = 10^5$Pa, $N_2$ gas.

Simulations showed that the evolution of the space charge distribution has several typical stages [see Fig. 2(c,d) and Fig. 3]. During the first stage ($t < 137$ps), the electric field formed by the ions' space charge is relatively small. The electrons emitted from the cathode and the electrons that are generated due to $N_2$ gas ionization in the cathode vicinity (at distances $\leq 14\mu$m) gain high energy from the external electric field and depart quickly from the cathode region, leaving ions. Therefore, the space charge distribution is characterized by two distinct regions; namely, the region with excess ions and the region with excess electrons.

The second stage of space charge distribution is obtained when the electric field of the accumulated ion space charge becomes sufficient to decelerate the RAE propagating toward the anode in the region OK$^*$ [see Fig. 2(a) and Fig. 3]. Also, this ion space charge increases the electric field in the cathode vicinity [region OK, Fig. 2(a)]. Thus, during this stage of the discharge one obtains three regions with different space charge distributions [see Fig. 3(a)]: the region with an excess of positive charge (KK*), the region of VC with an excess of negative charge (in the vicinity of K*), and the region K*A where the electron and ion densities are almost equal to each other.

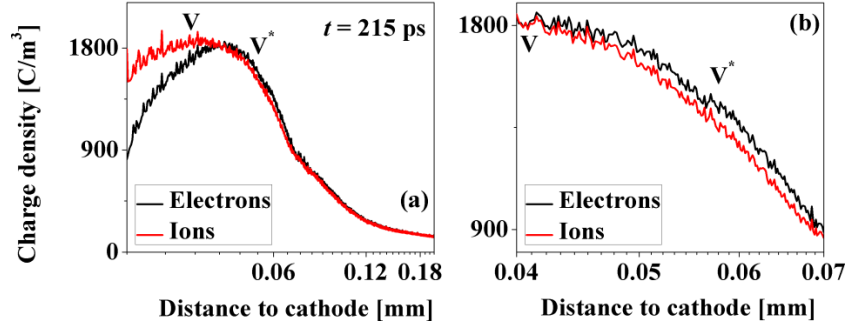

Fig. 3. (Color online) (a)-(c) Distributions of charge densities of electrons and ions at time of the virtual cathode formation; $T = 2$ns, $\varphi_0 = 120$kV, $P = 10^5$Pa, $N_2$ gas.

In fact, the formation of the potential hump and well [see Fig. 2(a)] influences on the initial stage of RAE generation significantly. Earlier ($t < 215$ps) in the accelerating pulse, the appearance of the potential hump due to ion accumulation increases the electric field at the cathode, which allows electrons emitted from the cathode to gain more energy than the electrons could gain in an undisturbed electric field. Later in the accelerating pulse, the depth of the potential well $\varphi_{OK*}$ is increased due to electron space charge accumulation at that location. The accumulated electrons in K* location have energies lower than the ionization potential of $N_2$ molecules and, therefore, these electrons cannot produce new electrons and ions. In addition, the considered amplitude and rise times of the voltage pulse are insufficient for ions to be shifted toward the K* location and to neutralize the VC electron space charge. The formation of potential well $\varphi_{OK*}$ leads to the cut-off and capture of low-energy electrons with $\varepsilon_e \leq e\varphi_{OK*}$ and the deceleration of electrons entering this well with $\varepsilon_e > e\varphi_{OK*}$. When the VC is formed, all the electrons accelerated in the KO region are captured in the KOK* region. Thus, the VC limits the generation of RAE formed of electrons emitted by the cathode and produced in the region KOK*.

The third stage of the discharge begins at $t > 260$ps and is characterized by the spreading of the VC and the appearance of sheaths with excessive negative or positive charges inside the VC at $r > 0.05$mm locations [see, Fig. 2(b,d)]. At the anode side of the VC one obtains electric field $E_{VC} > E_{cr}$, which allows continuation of the RAE generation from that location.

Fig. 4(a) shows a snapshot of the EEDF-1 calculated for electrons inside the CA gap for given times. Fig. 4(b) presents the EEDF-2 for electrons having $\varepsilon_e \geq 1$keV and reaching the anode within the time interval between the start of simulations and the considered time. One can see a significant spread in the EEDF-1 within the accelerating pulse with a maximum spectrum of electrons with $\varepsilon_e < 10$keV. Also, the EEDF-2 at the anode [see Fig. 4(b)] changes significantly within the accelerating pulse. Namely, at the beginning of the accelerating pulse ($t < 295$ps), EEDF-2 consists of electrons whose energy is within a relatively narrow range ($\Delta\varepsilon_e \approx 40$keV). Later in the accelerating pulse, when the cathode potential decreases ($t > 500$ps), EEDF-2 contains only a small part of the electrons with $\varepsilon_e > 50$keV, while the majority of the electrons has $\varepsilon_e < 30$keV.

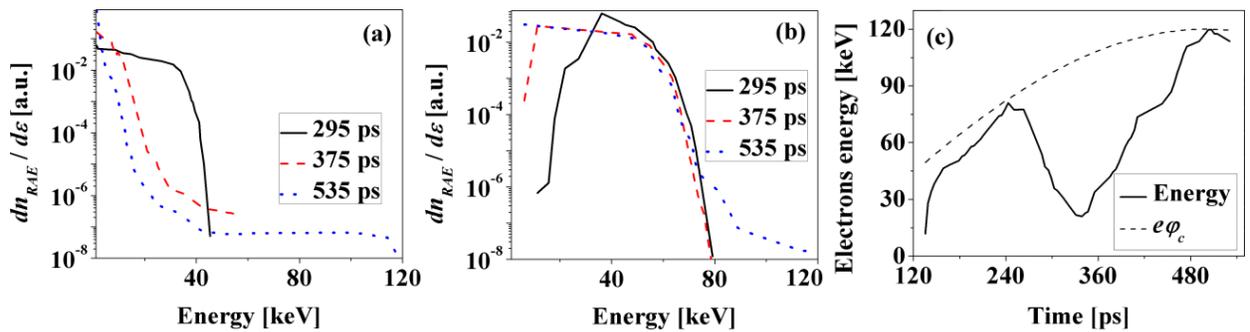

Fig. 4. (Color online) (a) Summarized EEDF in the cathode-anode gap at different times, (b) EEDF at the anode at different times, (c) the maximum electrons energy obtained in the CA gap; $T = 2$ns, $\varphi_0 = 120$kV, $P = 10^5$Pa, $N_2$ gas.

The time dependence of maximum electron energy obtained in the entire CA gap at different time points in the accelerating pulse is shown in Fig. 4(c). One can see that this dependence has two maxima. The first maximum corresponds to electrons with $\varepsilon_e \approx 81$keV, and it characterizes the RAE generated in the cathode vicinity that are cut off by the VC at the time of its formation. At $t \approx 240$ps, these electrons reach the anode. Within the time interval 240ps < $t$ < 360ps one obtains a decrease in the maximum electron energy inside the CA gap. The latter is explained by the potential well formation [see Fig. 2(a,b)], which decreases the energy of electrons entering this well from the cathode region and having $\varepsilon_e > e\varphi_{OK^*}$. The second maximum in electron energy, $\varepsilon_e \approx 118$keV, which is obtained at $t \approx 490$ps, corresponds to electrons that were accelerated between the VC and anode.

The presence of a gas in the diode changes significantly the physical processes that govern electrons generation. Fig. 5(a) shows the comparison between the numbers of electrons emitted per unit time, $dn_{em}/dt$, in the $N_2$-filled diode (here the space charge of emitted electrons and secondary electrons and ions was accounted for) and the vacuum diode (here the space charge of only emitted electrons was accounted for) and $dn_{em}/dt$ calculated using the FN law. One can see that the maximum $dn_{em}/dt$ in the $N_2$-filled diode is reached faster than in the vacuum diode. The latter can be explained by the opposite effect of secondary ions and electrons and emitted electrons' space charge on the electric field in the cathode vicinity. Namely, fast electrons leave secondary ions which cause the appearance of a non-compensated space charge that increases the electric field at the cathode [see Fig. 3 and Fig. 5(b)]. In the case of the vacuum diode, the $dn_{em}/dt$ value is limited only by the shielding of the electric field in the vicinity of the cathode by the emitted electrons' space charge. In addition, VC is not formed in the case of the studied parameters.

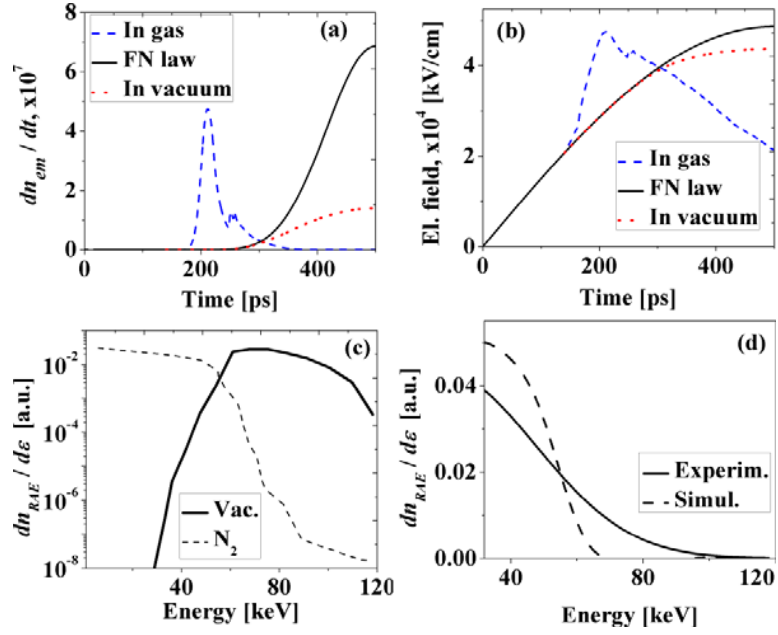

Fig. 5. (Color online) (a) Time dependence of $dn_{em}/dt$ in $N_2$-filled diode, in vacuum coaxial diode and calculated using Fowler-Nordheim law, (b) time dependence of electric field at the cathode surface for the same as in (a) diodes, (c) comparison between EEDF in $N_2$-filled diode and in vacuum coaxial diode at $t = 500$ps, (d) comparison of experimental results with results of simulations; $T = 2$ns, $\varphi_0 = 120$kV, $P = 10^5$Pa.

Fig. 5(c) shows the EEDF-2 at the anode at $t = 500$ps for the case of $N_2$-filled and vacuum coaxial diodes. One can see a drastic difference between these two spectra. Namely, the EEDF-2 maximum in the $N_2$-filled diode corresponds to an energy range $\varepsilon_e < 50$keV and in the case of the vacuum diode, the EEDF-2 maximum corresponds to electrons with $\varepsilon_e > 60$keV, i.e., electron energy follows $e\varphi_c$. The electron energy spectrum in the $N_2$-filled diode can be explained by the formation of the potential well [see, Fig. 2(a,b)], which decreases the energy of electrons penetrating this well toward the anode, and by the electron's energy losses in inelastic collisions. The excess of the electric field at the cathode above its vacuum value [see Fig. 4(b)], caused by the accumulated ion space charge, leads to a fast increase in the electron emission [see Fig. 4(a)].

Also, a comparison between the experimentally obtained EEDF (for details see in Ref. 25) and simulated EEDF-2 is presented in Fig. 5(d). The experimental spectrum was measured in the air ($P = 10^5$Pa) for $\varphi_0 \approx 120$kV, voltage rise time ~ 0.5ns, and $d_{ca} = 1$cm. One can see a satisfactory agreement between the calculated and experimental spectra. The slightly low energy in the simulated spectrum could be caused by a difference between the experimental and simulation HV pulse waveforms and diode configurations. Also, the simulations did not take into account collisions between electrons and oxygen molecules, which could influence the spectrum. Nevertheless, both the simulation and experimental spectra show that only a small part of the electrons has energies $\varepsilon_e \sim e\varphi_c$, while the main majority has energies $\varepsilon_e < 40$keV.

Time dependences of emitted electrons ($dn_{em}/dt$) at different values of $\varphi_0$ for the $N_2$-filled diode are shown in Fig. 6(a). One can see that the maximal value of $dn_{em}/dt$ does not depend on the value of $\varphi_0$. The increase in value of $\varphi_0$ leads only to reaching $dn_{em}/dt$ of its maximal value earlier. Also, one can see temporal oscillations in $dn_{em}/dt$. The simulation results showed that earlier in the accelerating pulse ($t < 150$ps), when the accumulated ion and secondary and emitted electrons space charges are rather small, the value of $dn_{em}/dt$ is determined by the external electric field. Later in the accelerating pulse, when the electric field at the cathode increases due to accumulated ion space charge, the value of $dn_{em}/dt$ increases faster than in the case of a non-disturbed electric field [see Fig. 5(a)]. This increase in $dn_{em}/dt$ depends exponentially on the electric field following FN law[19] and it continues till the emitted electrons' space charge becomes sufficient to screen electric field in the cathode vicinity.

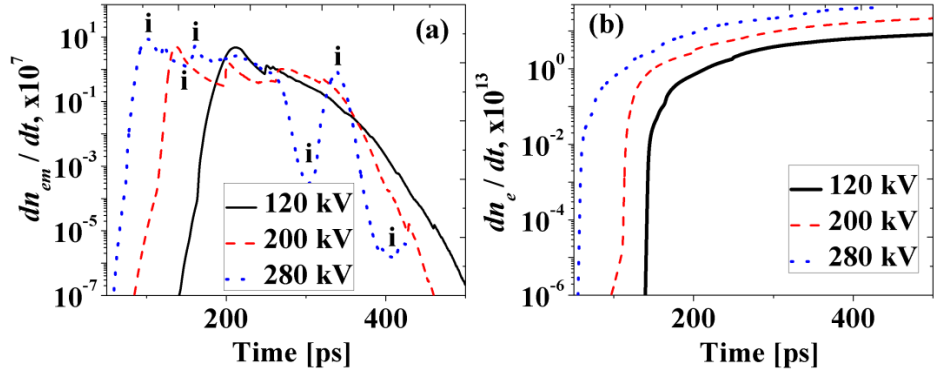

Fig. 6. (Color online) Time dependences of (a) emitted $dn_{em}/dt$ and (b) total $dn_e/dt$ at different amplitudes of the cathode potential; $T = 2$ns and $P = 10^5$Pa, $N_2$ gas.

This is the reason why one obtains almost equal $dn_{em}/dt$ for different values of $\varphi_0$. The obtained oscillations in $dn_{em}/dt$ values are related to the temporal evolution of electron and ion space charges in the cathode vicinity. Namely, the density of generated ions continues to be negligibly small in the cathode vicinity (< 1μm) and, therefore, the shielding effect [see first left point "i" in Fig. 6(a)] decreases $dn_{em}/dt$ and, respectively, limits RAE generation. However, the electric field at the cathode is determined also by the space charge of the generated ions at larger distances from the cathode. Therefore, at the time when the sum of the external electric field and electric field produced by ion space charge becomes larger than the electric field of the emitted electrons space charge, one obtains once more an increase in $dn_{em}/dt$ [see points denoted as "i" in Fig. 6(a)]. This increase is terminated when the screening effect becomes dominant once more. Thus, one obtains oscillations in the value of $dn_{em}/dt$.

The total number of electrons $dn_e/dt$, i.e., electrons which one obtains inside the CA gap at a given time, increases exponentially [~ $exp(\alpha t)$] with two typical stages having different powers $\alpha$ in their exponents [see Fig. 6(b)]. The first stage, characterized by intensive ionization of gas by accelerated electrons, has values of $\alpha$: $9.9 \cdot 10^{11} s^{-1}$, $1.2 \cdot 10^{12} s^{-1}$, and $1.7 \cdot 10^{12} s^{-1}$ for 120kV, 200kV and 280kV, respectively. During the second stage, when the shielding effect becomes significant, one obtains a decrease in the values of $\alpha$: $0.6\cdot10^{10}$s$^{-1}$, $1.3\cdot10^{10}$s$^{-1}$ and $1.9\cdot10^{10}$s$^{-1}$ for 120kV, 200kV and 280kV, respectively.

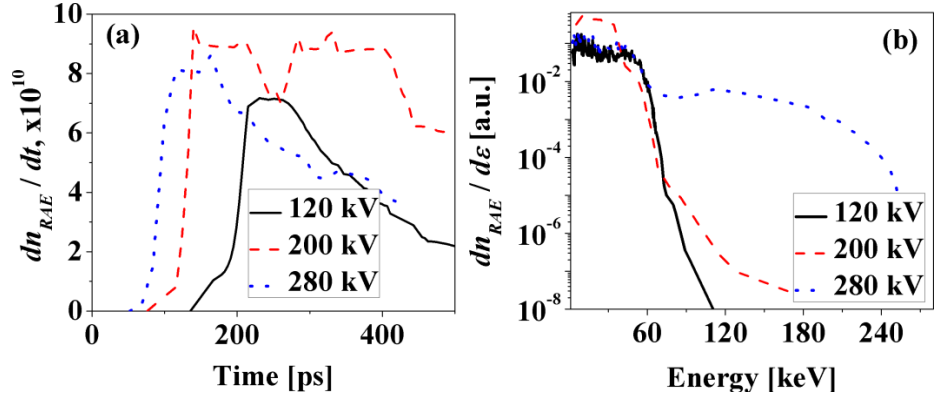

Fig. 7. (Color online) (a) Time dependence of RAE $dn_e/dt$ (total number of electrons with $\varepsilon_e \geq$ 1keV) at different amplitudes of cathode potential, (b) EEDF-2 at the anode for different amplitudes of the cathode potential at $t = 500$ps; $T = 2$ns, $P = 10^5$Pa, N$_2$ gas.

A comparison of the results presented in Fig. 6(a) and Fig. 7(a) shows that RAE generation continues also after the shielding effect limits the electron emission from the cathode. Also, it was obtained that independently of the $\varphi_0$, the total amount of emitted electrons prior to the VC formation is $\sim 10^9$. At the same time the largest RAE (i.e., electrons with $\varepsilon_e > 1$keV) quantity in the CA gap reaches $\sim 10^{11}$ [Fig. 7(a)]. Therefore, one can conclude that the majority of the RAE are the secondary electrons generated and accelerated in the CA gap.

The results of simulations for $T = 2$ns showed that the increase in the value of $\varphi_0$ decreases the time $t_{VC}$ of VC formation ($t_{VC} \approx 215$ps for $\varphi_0 = 120$kV, $t_{VC} \approx 138$ps for $\varphi_0 = 200$kV, and $t_{VC} \approx 103$ps for $\varphi_0 = 280$kV) due to more intense electron emission. At that time, the VC potential is $\varphi_{VC} \approx 75$kV for $\varphi_0 = 120$kV, $\varphi_{VC} \approx 84$kV for $\varphi_0 = 200$kV, and $\varphi_{VC} \approx 89$kV for $\varphi_0 = 280$kV. The earlier formation of the VC leads to an earlier termination of the generation of RAE composed of electrons emitted from the cathode and electrons formed as a result of ionization inside the cathode-VC gap [see Fig. 7(a)].

The simulations showed that the VC is formed prior to the first RAE reaching the anode. Thus, the maximal energy of electrons that have passed the location of the potential well where the VC will be formed and reach the anode is ~ 75keV, ~ 84keV and ~ 89keV, for value of $\varphi_0$ 120kV, 200kV, and 280kV, respectively. Also, the results of the simulations showed that the maximum value of $E_{VC}$ toward the anode at the moment of the VC formation is $E_{VC} < E_{cr}$ and, therefore, VC cannot be considered as a source of RAE. As a result, the RAE amount is almost constant during the next $\Delta t \sim $ 20ps [see Fig. 7(a)]. This time interval is the time-of-flight of RAE existing inside the VC-anode gap toward the anode. Later in the accelerating pulse, simulations showed $E_{VC} > E_{cr}$ and, respectively, the persistence of RAE generation by electrons emitted from the VC location. The decrease in the amount of the RAE at that time is related to the flow of these electrons through the anode and the smaller amount of electrons emitted by the VC than in the case of electrons emission from the cathode.

Calculated spectra of RAE that reached the anode are shown in Fig. 7(b) for different $\varphi_0$ at $t$ = 500ps. One can see that the maximal electron energy at $\varphi_0$ = 120kV is $\varepsilon_e \approx$ 118keV, although at $\varphi_0$ = 200kV and $\varphi_0$ = 280kV one obtains $\varepsilon_e \approx$ 150keV and $\varepsilon_e \approx$ 250keV, respectively. The difference between the maximal electron energy and $e\varphi_0$ is explained by the different potential distributions in the CA gap and the different time dependences of the emitted $dn_{em}/dt$ for different $\varphi_0$. For instance, Fig. 6(a) shows that at 500ps, when the cathode potential reaches $\varphi_0$, FE is weak at $\varphi_0$ = 200kV and $\varphi_0$ = 280kV but FE produces several electrons per time step for $\varphi_0$ = 120kV.

A decrease in the rise time of the accelerating voltage leads also to a decrease in the value of $t_{VC}$ and an increase in the value of $\varphi_{VC}$. For instance, for $\varphi_0$ = 120kV, $d_{ca}$ = 1cm and $P = 10^5$Pa one obtains $t_{VC} \approx$ 68ps and $\varphi_{VC} \approx$ 91kV for $T$ = 0.5ns, $t_{VC} \approx$ 123ps and $\varphi_{VC} \approx$ 84kV for $T$ = 1.0ns and $t_{VC} \approx 215$ps and $\varphi_{VC} \approx 75$kV for $T = 2.0$ns. Thus, one can conclude that the shorter the rise time of the accelerating voltage, the faster is the rate of RAE generation [see Fig. 8(a)].

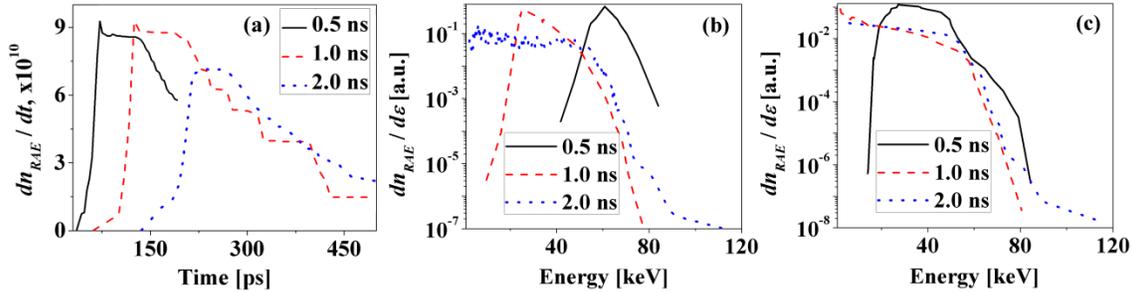

Fig. 8. (Color online) (a) Time dependence of the RAE $dn_{RAE} / dt$ for different values of $T$, (b) integrated EEDF at the anode for different values of $T$ at time $T/4$, (c) integrated EEDF at the anode for different values of $T$ at time $T/2$; $\varphi_0 = 120$kV, $P = 10^5$Pa, $N_2$ gas.

Results of EEDF simulations at the anode at $t = T/4$ for different values of $T$ are shown in Fig. 8(b). One can see that the longer the rise time of the accelerating voltage, the low-energy and broader is the obtained EEDF. Also, the simulations showed that the amount of RAE reaching the anode at $t = T/4$ decreases with the decrease in the accelerating voltage rise time. These RAE are the electrons that are generated prior to the VC formation. EEDF at the anode obtained during the entire accelerating pulse, i.e., at $t = T/2$, are shown in Fig. 8(c). Similarly to the EEDF obtained at $t = T/4$, an increase in pulse duration leads to a broadening of the energy spectrum and its shift to a low-energy range. Here, let us note that these spectra include electrons that are generated before and after the VC formation.

Simulations have shown that the process of the VC formation depends on the value of $\varphi_0$ and gas pressure. For instance, the VC formation is not obtained at $\varphi_0 = 60$kV, $T = 1$ns and $P = 10^5$Pa and $P = 2\times10^5$Pa, but the VC is formed at $\varphi_0 = 60$kV and $P = 3\times10^5$Pa. Fig. 9(a) shows potential distributions at different pressures at $t = 250$ps, i.e., when the cathode potential reaches its maximum value of 60kV. One can see that the VC is formed only at $P = 3\times10^5$Pa at a distance $r \approx 0.04$mm at $t_{VC} \approx 220$ps and its location shifts toward the anode [Fig. 9(b)]. Here let us note that the maximal value is $E_{VC} < E_{cr}$ and, therefore, the VC cannot be considered as a RAE source. The lack of VC formation at lower pressures results in RAE generation with broader spectra than at $P = 3 \times 10^5$Pa. Nevertheless, the energy interval corresponding to the EEDF maximum (16-18keV) weakly depends on pressure [see Fig. 9(c)].

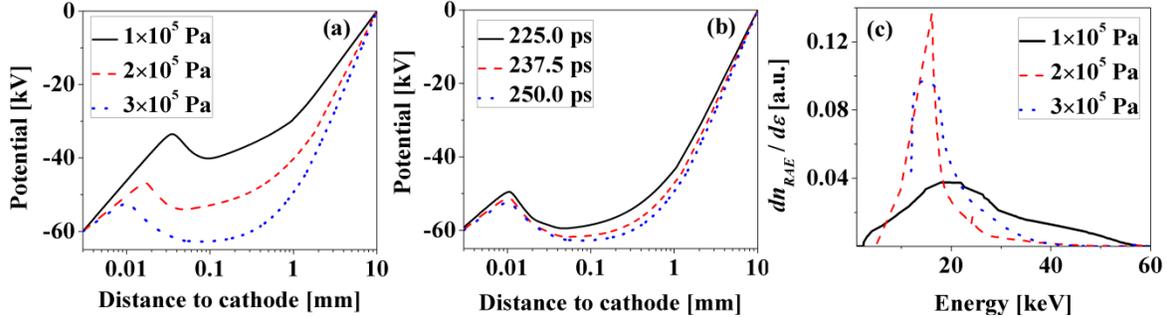

Fig. 9. (Color online) (a) Potential distribution at different gas pressures at $t = 250$ps, (b) potential distribution for $P = 3 \times 10^5$Pa, (c) EEDF at the anode for different $N_2$ gas pressure at $t = 250$ps; $\varphi_0 = 60$kV, $T = 1$ns.

In order to compare the process of RAE generation in different gases, simulations were carried out for He gas for $P = 10^5$Pa, $T = 2.0$ns and $\varphi_0 = 120$kV. For ionization cross-sections the NIST database[15] was used and cross-sections of elastic scattering and excitation of the first electron's level of He were extrapolated for the high-energy range (up to 200keV) using the data presented in Ref. 26. The maximum values of ionization cross-sections for $N_2$ and He are obtained at almost the same $\varepsilon_e \sim 115$eV. However, the ionization cross-section for $N_2$ is $\sim 10$ times larger than that for He.[16] In addition, the ionization energy for $N_2$ ($\varepsilon_{ion} = 15.6$eV) is smaller than for He ($\varepsilon_{ion} = 22.5$eV) and, therefore, the value of $E_{cr}$ in He at $P = 10^5$Pa is $E_{cr} = 1.14 \times 10^5$V/cm, i.e., 3.9 times lower than the value of $E_{cr}$ in $N_2$ gas. Therefore, the RAE generation is more efficient in He than in $N_2$ for the same pressure. Indeed, numerical simulations showed that the RAE current amplitude and pulse duration are significantly larger in He gas than in $N_2$ gas. Also, it was shown that the ratio between the total amount of RAE and emitted electrons is larger in He than in $N_2$ [see Fig. 6(a), Fig. 7(a) and Fig. 10(a)].

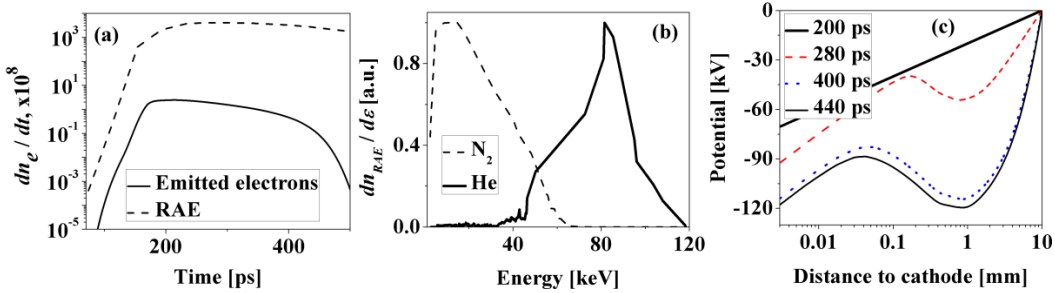

Fig. 10. (Color online) (a) Time dependence of emitted $dn_{em}/dt$ and RAE $dn_{RAE}/dt$ in He as working gas, (b) comparison between EEDF on anode in He in $N_2$ at 500ps, (c) potential distribution in He at different times; $T = 2$ns, $\varphi_0 = 120$kV, $P = 10^5$Pa.

The comparison between the calculated EEDF-2 of RAE in $N_2$ and He at $T = 2$ns, $\varphi_0 = 120$kV, and $P = 10^5$Pa is shown in Fig. 10(b). One can see that the EEDF in He gas is characterized by maximum at $\varepsilon_e \approx 91$keV, though the maximum of EEDF in $N_2$ gas corresponds to $\varepsilon_e \approx 8$keV. Such large difference in spectra can be explained by a few factors. The first factor is different electron energy losses in $N_2$ and He gases: they are significantly smaller in He gas. The second factor is the different times of the VC formation. Indeed, in the case of $N_2$ gas, the VC is formed at $t \approx 215$ps when $\varphi_C \approx 75$ kV and for He gas, the VC is formed at $t \approx 400$ps when $\varphi_C \approx 114$kV. The latter leads to a larger amount of high-energy electrons than one obtains in EEDF for the case of He gas. In addition, in the case of $N_2$ gas, $E_{VC} < E_{cr}$ and in the case of He gas, $E_{VC} \approx 1.32 \times 10^5$V/cm, which is larger than the $E_{cr}$ value for He gas. Thus, in the case of He gas, after the VC formation, the latter becomes a source of RAE. Finally, it is interesting that the VC position in He gas [see Fig. 9(c)] is almost immovable as compared with that in $N_2$ gas [see Fig. 3(b)].

**IV. SUMMARY**

The RAE generation in pressurized $N_2$ and He gases was investigated using 1D PIC numerical code for coaxial diode geometry. These simulations included the potential distribution evolution caused by the space charge of generated electrons and ions and the dependence of the electron field emission on the electric field at the cathode.

It was shown that the RAE generation occurs in two stages. The first stage continues until the VC is formed. RAE generated during this stage are composed of electrons emitted from the cathode and generated in the cathode vicinity. The second stage occurs when the VC formation has happened. The latter terminates RAE generation till the electric field at the VC becomes larger than $E_{cr}$, which allows one to continue the generation of RAE emitted from the VC location. Simulations have shown that the first stage contributes to the RAE generation more significantly than the second stage.

Simulations showed that the shielding of the FE of electrons at the initial stage of the pressurized gas discharge plays a major role in the generation of RAE. This shielding occurs by the space charge of generated secondary electrons and ions and emitted electrons and it leads to a significant change in the space- and time evolution of the electric field distribution in the CA gap. Since the VC formation and shielding effect were obtained during the voltage rise time, RAE generation also started and terminated at that time interval.

An analysis of the processes that accompany RAE generation has shown that the parameters of RAE depend strongly on gas type and its pressure, cathode potential, and voltage rise time. An increase in the cathode potential decreases the time of the VC formation which leads to faster termination of RAE generation. The simulated and experimentally obtained EEDF showed satisfactory agreement. Namely, the obtained electron energy distributions on the anode contained electrons in a wide energy range, with the majority of RAE having energies much smaller than $e\varphi_c$. Also, it was shown that the shorter the rise time of the accelerating voltage, the faster is the rate of the RAE generation and a more energetic EEDF is obtained. Finally, comparison of RAE generation in $N_2$ and He gases has shown that the amplitude and duration of the RAE current are larger in He as compared with $N_2$ gas due to later formation of the VC in the case of He gas.

## ACKNOWLEDGEMENTS

This work was supported in part at the Technion by a fellowship from the Lady Davis Foundation, the Technion grant #2013371 and the Center of Absorption in Science, Ministry of Immigrant Absorption, State of Israel.